\documentclass{article}

\usepackage{arxiv}
\usepackage{amsmath}
\usepackage[utf8]{inputenc} 
\usepackage[T1]{fontenc}    
\usepackage{hyperref}       
\usepackage{url}            
\usepackage{booktabs}       
\usepackage{amsfonts}       
\usepackage{nicefrac}       
\usepackage{microtype}      
\usepackage{lipsum}
\usepackage{graphicx}
\usepackage{xcolor}
\usepackage{verbatim}
\graphicspath{ {./images/} }

\title{Statistical inference of lead-lag at various timescales between asynchronous time series from p-values of transfer entropy }

\author{
 Christian Bongiorno \&
Damien Challet \\
Université Paris-Saclay, CentraleSupélec\\
Laboratoire de Mathématiques et Informatique pour la Complexité et les Systèmes\\
  91192 Gif-sur-Yvette, France

}

\begin{document}
\maketitle

\begin{abstract}
    Symbolic transfer entropy is a powerful non-parametric tool to detect lead-lag between time series. Because a closed expression of the distribution of Transfer Entropy is not known for finite-size samples, statistical testing is often performed with bootstraps whose slowness prevents the inference of large lead-lag networks between long time series. On the other hand, the asymptotic distribution of Transfer Entropy between two time series is known. In this work, we derive the asymptotic distribution of the test for one time series having a larger Transfer Entropy than another one on a target time series. We then  measure the convergence speed of both tests in the small sample size limits via benchmarks. We then introduce Transfer Entropy between time-shifted time series, which allows to measure the timescale at which information transfer is maximal and vanishes. We finally apply these methods to tick-by-tick price changes of several hundreds of stocks, yielding non-trivial statistically validated networks. 
\end{abstract}


\section{Introduction}

Causality is the process from which an event influences the outcome of another process. In real-world data, detecting causal links may be hard. An intuitive way to infer causality consists of measuring lagged correlation between the states of two variables. Although this approach is in line with what is humanly perceived as a cause-effect relationship, these lagged correlations may, in many cases, come from the influences of hidden variables. Furthermore, observing significant lagged correlations is not a necessary condition for the inference of a cause-effect relationship because such a measure can seem negligible in a highly non-linear and noisy system.

One of the early approaches to cause-effect detection is Granger causality~\cite{granger1969investigating}. The idea of Granger causality is to assume that \textit{B causes A} is equivalent to \textit{B forecasts A}. Forecasting at that time was typically addressed with auto-regressive models. Therefore, Granger causality is able to detect causality only when a linear model is a good approximation to the dynamics of $A$ and $B$. Needless to say, most real-world systems are not linear.

More generally, $A$ can only be predicted from $B$ if a significant amount of information is transferred from $B$ to $A$. 
Transfer Entropy (TE)~\cite{schreiber2000measuring} measures the information gain from the past of $B$ in predicting the future of $A$ after discounting of self-information from the past of $A$. TE is more general than Granger causality. Both are equivalent for auto-regressive models with Gaussian variables~\cite{barnett2009granger}. While TE can be defined for continuous non-linear models, it becomes model-free when the states of  $A$ and $B$ take or are approximated by discrete values. If the original models are not genuinely discrete, the longer the time series, the more refined the discrete description of the original time series can be. This is known as symbolic TE~\cite{staniek2008symbolic}, which provides an unbiased estimator of the model-free TE. However, when only finite time series are available,  the error of sample probabilities and thus of the sample TE increases with the chosen number of distinct discrete values. Such an irreducible error makes it difficult to distinguish a real signal from random noise fluctuations (for more spurious sources of TE, see \cite{smirnov2013spurious}).

Testing the significance of TE can be done with data shuffling or resampling data, which is slow. Direct testing was proposed in Ref.~\cite{barnett2012transfer}, which introduced a null model for TE and shows that TE asymptotically converges to a chi-square distribution with 2 degrees of freedom for binary time series. A generalization of this test to more generic discrete time series for Markovian systems is found in Ref.~\cite{kontoyiannis2016estimating}. Finally, TE between asynchronous time series was recently defined in Ref.~\cite{harre2015entropy}.

Our main contributions are as follows: first, we derive an asymptotic statistical test to compare the amount of TE from $B$ to $A$ and from $C$ to $A$, which allows us to significantly rank predictors pairs based on their respective information content. Secondly, we quantify the finite-size effects of the asymptotic tests by comparing their outcomes with those of bootstrap tests. Thirdly,  we use time-shifting of the target time series in an asynchronous setting so as to infer significant time delays and the delay after which no statistically significant TE or TE difference takes place. Third, we demonstrate the effectiveness of our methods by inferring a statistically validated lead-lag network of tick-by-tick price changes from a large database of financial data and by characterising how TE changes as a function of time delays.

\section{Significance test for Transfer Entropy}\label{sec:tepv}

Let us fix some notations. For any time $t$, information about the past of time series $A$ is encoded by $n_A$ different states. The simplest case is binary information (two symbols) about some previous time step $t'< t$; binary information about $M$ different time steps amounts to $n_A=2^M$ possible states. Note that the number of possible symbols per time step may change for each $t'$ in a fixed way. The number of possible states that describe the future of $A$ is denoted by $n_A^+$. Similarly, the number of states that describe time series $B$ at time $t$ is $n_B$.

The transfer entropy from $B$ to $A$ measures the additional information provided by $B$ with respect to the information about the past of $A$  when one tries to predict $A$. It is defined by the difference between the conditional entropy $H(A^+|A)$ of the future of $A$, denoted by $A^+$, on its past, and the entropy $H(A^+|A,B)$ of $A^+$ conditional  on the past of $A$ and $B$, as follows
\begin{equation}
\mbox{TE}^{B \to A} = H(A^+|A)- H(A^+|A,B).
\end{equation}

\begin{figure}
\centering
\includegraphics[width=0.45\columnwidth]{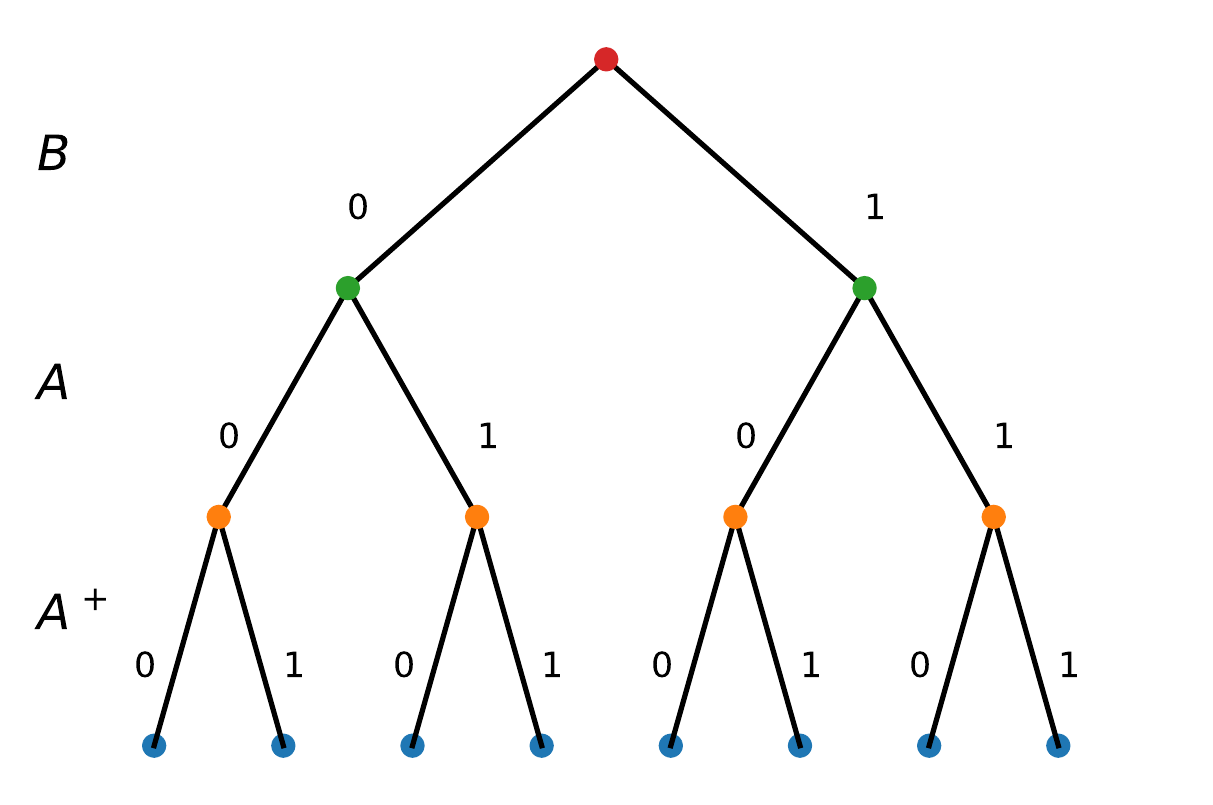}
\includegraphics[width=0.45\columnwidth]{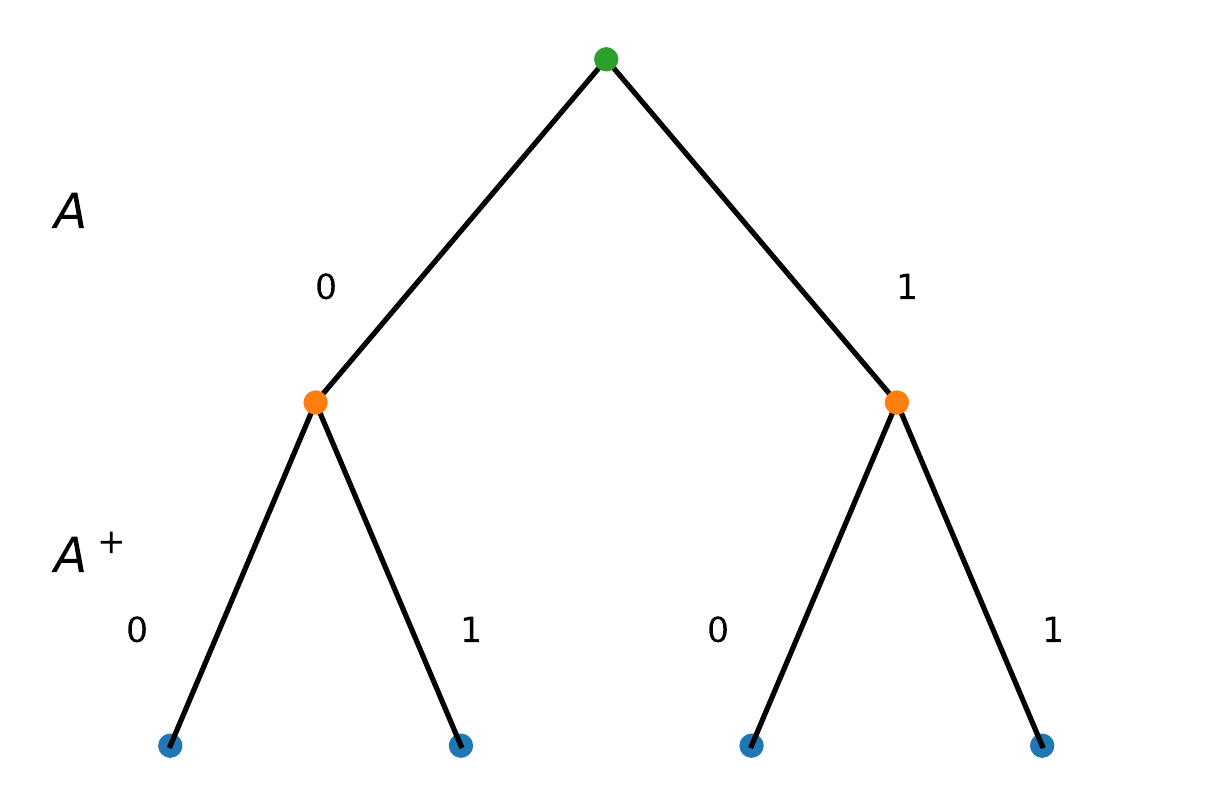}
\caption{Example of the maximum likelihood binary tree (on the left) that includes $B$, and the alternative model (on the right) that does not include $B$. \label{fig:DTMs}}
\end{figure}

Refs.\ \cite{barnett2012transfer,kontoyiannis2016estimating} show that the $\mbox{TE}^{B\to A}$ converges to a chi-square distribution for binary states and Markovian systems with constant numbers of symbols per time step. As a warm-up for the next section, let us derive this result from the comparison between two decision tree models (see Fig.\ \ref{fig:DTMs}).

When testing for the presence of significant transfer entropy from $B$  to $A$, the null hypothesis $\mathcal{H}_0$ and the alternative hypothesis $\mathcal{H}_1$ are respectively
\begin{eqnarray}
  \mathcal{H}_0: \mbox{TE}^{B \to A}=0\\
  \mathcal{H}_1: \mbox{TE}^{B \to A}>0.
\end{eqnarray}

The likelihood of the decision tree model that includes the series $B$ equals
\begin{equation}
    \log \mathcal{L}_{B A A^+} = T \sum_{b=1}^{n^B} \sum_{a=1}^{n^A} \sum_{a^+=1}^{n^{A^+}}  \mathcal{P}(a^+,a,b)  \log \mathcal{P}(a^+|a,b) = - T \cdot  H(A^+|A,B),
\end{equation}
where $\mathcal{P}(a^+|a,b)$ and $\mathcal{P}(a^+,a,b)$ are the sample probabilities.

The likelihood of the decision tree model that does not include the series $B$ is given by
\begin{equation}
    \log \mathcal{L}_{A A^+} = T  \sum_{a=1}^{n^A} \sum_{a^+=1}^{n^{A^+}}  \mathcal{P}(a^+,a)  \log \mathcal{P}(a^+|a) = - T \cdot  H(A^+|A).
\end{equation}

\begin{figure}
\centering

\includegraphics[width=0.45\columnwidth]{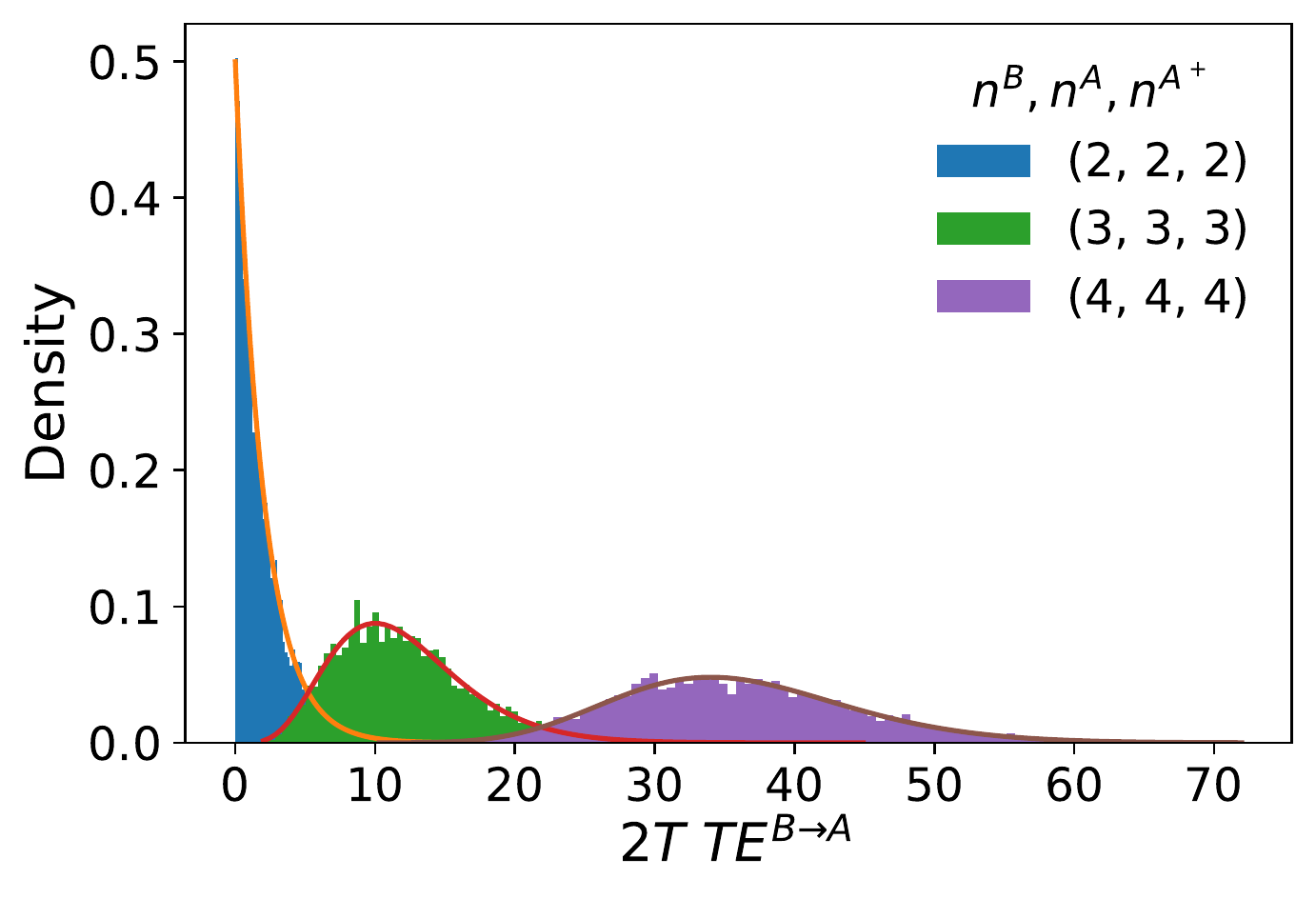}
\includegraphics[width=0.45\columnwidth]{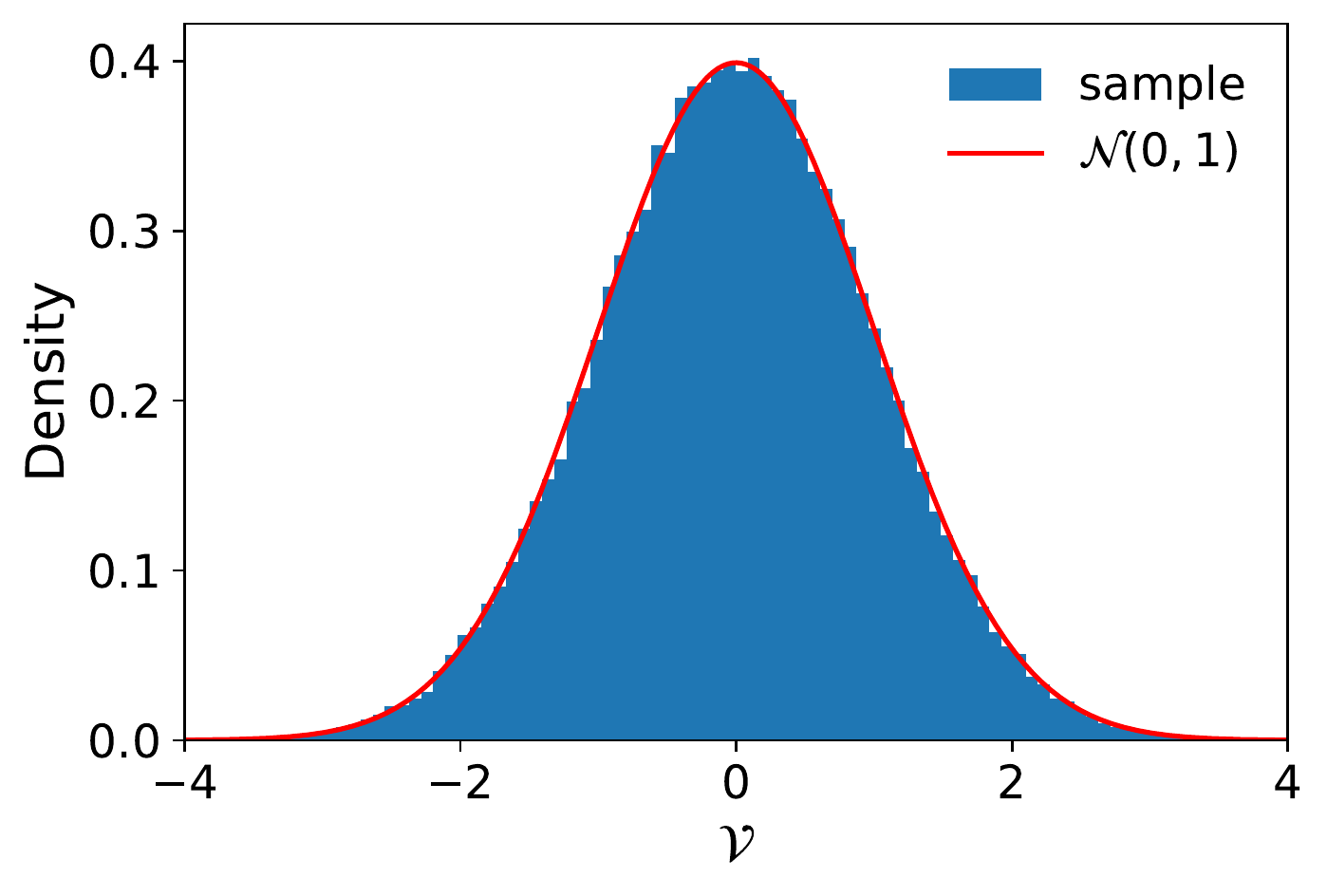}

\caption{The left panel shows the distribution of $2T\cdot \mbox{TE}^{B \to A}$ obtained with reshuffling of the data-series $B$ for different values of the number of symbols. The lines are the theoretical expectation from the $\mathcal{X}^2(\nu)$ distribution. The right panel Distribution of the Vuong metric obtained with independent 100,000 realizations of a time series of T=10,000 records for $S=0.2$.\label{fig:TEnull} }
\end{figure}


If the null hypothesis is true, the knowledge of $B$ does not improves significantly  predictability and thus the difference between both likelihoods will be due to over-fitting of the sampling error. In such a case, Wilks' theorem \cite{wilks1938large} states that
\begin{eqnarray}
   2 (\log \mathcal{L}_{B A A^+} - \log \mathcal{L}_{A A^+}) = -2 T \left( H(A^+|A,B) - H(A^+|A) \right) = \nonumber \\
   = 2 T \cdot \mbox{TE}^{B \to A} \xrightarrow[]{d} \mathcal{X}^2\left( \nu \right)
\end{eqnarray}\label{eq:like_TE}
  
converges to a chi-square distribution in the limit of large sample size with $\nu$ degrees of freedom given by
\begin{equation}\label{eq:df1}
    \nu = n^A \left( n^{A^+} -1 \right) \left(n^B-1\right).
\end{equation}
This number can be obtained by counting the difference in the branches of the two decision trees. Specifically the number of degrees of freedom of the model $\mathcal{P}(a^+| a,b)$ is $\nu^{AB} = n^A n^B (n^{A^+} -1)$ where $\sum_{a^+=1}^{n^{A^+}} \mathcal{P}(a^+| a,b) = 1$, whereas from the model $\mathcal{P}(a^+| a)$ one finds $\nu^{A} = n^A (n^{A^+} -1)$. Given that  $\nu = \nu^{AB} - \nu^{A}$,  \eqref{eq:df1} follows. If the number of symbols per time step is constant and if the maximum number of relevant time steps in the past is also constant, one recovers the result of Ref.\ \cite{hadash2018estimate}. Figure \ref{fig:TEnull} (left) compares the distributions given by Eq.\ \eqref{eq:like_TE} and numerical simulations for various parameter values.

\section{Comparing Two Transfer Entropy measures}\label{sec:over}

The same approach can be used to test if time series $B$ is more informative than $C$ to predict $A$. In other words, we aim to test if the real underlying TEs are such that
\begin{eqnarray}
   \mathcal{H}_0 :\mbox{TE}^{B \to A}=\mbox{TE}^{C \to A} \\ 
   \mathcal{H}_1 :\mbox{TE}^{B \to A}> \mbox{TE}^{C \to A}. 
\end{eqnarray}

Let us introduce the difference between the estimated TEs
\begin{equation}
    Q = \mbox{TE}^{B \to A} - \mbox{TE}^{C \to A} = H(A^+|A,C) - H(A^+|A,B).
\end{equation}
The likelihood of the model with $B$ is 
\begin{equation}
       \log \mathcal{L}_{B A A^+} = T \sum_{x=1}^{n^B} \sum_{a=1}^{n^A} \sum_{a^+=1}^{n^{A^+}}  \mathcal{P}(a^+,a,b)  \log \mathcal{P}(a^+|a,b) = - T \cdot  H(A^+|A,B) 
\end{equation}
while the likelihood of the model with $C$ is 
\begin{equation}
       \log \mathcal{L}_{C A A^+} = T \sum_{c=1}^{n^C} \sum_{a=1}^{n^A} \sum_{a^+=1}^{n^{A^+}}  \mathcal{P}(a^+,a,c)  \log \mathcal{P}(a^+|a,c) = - T \cdot  H(A^+|A,C).
\end{equation}
The difference in likelihood is therefore
\begin{equation}
      \Delta \mathcal{L} = \log \mathcal{L}_{BAA^+} -  \log  \mathcal{L}_{CAA^+} = T \cdot Q.
\end{equation}

To obtain the distribution of the difference in likelihood of the maximum likelihood models, let us call $\omega$ the standard deviation of the point-wise likelihood difference
\begin{equation}
    \omega^2 = \mbox{Var}\left[ \log \mathcal{P}(a^+| a, b) -\log \mathcal{P}(a^+| a, c)  \right]
\end{equation}
computed over the data-series of length T of elements $(b,c,a,a^+)$.

Then for Vuong's closeness test\ \cite{vuong1989likelihood} states that, if the null hypothesis is true,
\begin{equation}
    \mathcal{V}:=\frac{\Delta \mathcal{L} -\nu}{\sqrt{T} \omega} \xrightarrow[]{d} \mathcal{N}(0,1),
\end{equation}
where $\nu=E(\Delta \mathcal{L})$, converges to a standardized normal distribution for $T\to\infty$, see Fig.~\ref{fig:TEnull} (right).

In fact, $\nu$ is the difference between the number of degrees of freedom of the two trees associated with $B\to A$ and $C\to A$. While the estimation of $\nu$ from the trees is hard for overlapping models, but given that the hypothesis of Vuong's test applies, it is possible to estimate it as follows: if the null hypothesis is true, $\mbox{TE}^{B \to A}=\mbox{TE}^{C \to A} = S > 0$ and  $\Delta \mathcal{L}$ is the difference between two non-central chi distribution~\cite{barnett2012transfer}, divided by $2$, with expected value
\begin{eqnarray}
  E[T\, \mbox{TE}^{C \to A}] = \frac{1}{2}\left( \nu^C + S\right) =\frac{1}{2}\left[ n^A \left( n^{A^+} -1 \right) \left(n^C-1\right)+ S\right]  \\
  E[T\, \mbox{TE}^{B \to A}] = \frac{1}{2}\left( \nu^B + S\right) =\frac{1}{2}\left[ n^A \left( n^{A^+} -1 \right) \left(n^B-1\right)+ S\right].  
\end{eqnarray}
The degrees of freedom $\nu^C$ and $\nu^B$ are given by\ \eqref{eq:df1}. Therefore $E[\Delta \mathcal{L}- \nu]=0$ if
\begin{equation}
    \nu = (n^B - n^C) \frac{n^A}{2} (n^{A^+}-1).
\end{equation}

\subsection{Bootstrap p-value}
Let us briefly describe two generic approaches to associate a p-value to TE from bootstrap resampling, which can of course be computationally intense. For example,    significant p-values corrected by  multiple comparison techniques  must be orders of magnitude smaller than the intended significance threshold, which requires a very large number of bootstrap samples. 
For both approaches we must identify an event event matrix $E \in  \mathbb{R}^{T \times 3}$ where $e_t = (a'_t, a_t, b_t)$ is a generic event. This  matrix can be  generalized to more than two variables. 

\subsubsection{Significant TE via shuffling}
To estimate the p-value of sample $\mbox{TE}^{B\to A}=S$, it is enough to perform a permutation test of $B$, so as to break any link between the past of $B$ and the future of $A$. This kind of approach requires many  reshuffling of column $B$ of matrix $E$. Let us denote by  $E_{(r)}$ the data matrix $E$ whose $B$ column has been reshuffled, $r$ being the index of a particular permutation of column $B$. This new matrix preserves both $H(A^+|A)$ and $H(B)$ but breaks any information flow between $B$ and $A$. From $E_{(r)}$, one computes $\mbox{TE}_{(r)}^{B \to A}=S_{(r)}$. By performing $q$ independent re-shuffling and TE estimations, the p-value can be defined as
\begin{equation}
    p\mbox{-value} = \frac{\sum_{i=1}^q\delta(S>S_{(i)})}{q},
\end{equation}
where $\delta(x)=1$ if the assertion is true and zero otherwise.
\subsubsection{Comparing TEs via bootstraps}
Assume that both $\mbox{TE}^{B \to A}=S^{B}$ and $\mbox{TE}^{C \to A}=S^{B}$ are significantly larger than zero. A different bootstrap procedure must be adopted. In this case, indeed, the bootstrap is used to associate a confidence interval to $S^B$ and $S^C$. Let us start by defining the event matrix $E$ with elements $e_t = (a'_t, a_t, b_t, c_t)$, then a bootstrap realization of $E_{(r)}$ is obtained by re-sampling with replacement the events such that the bootstrap event matrix $E_{(r)}$ has the same sample size of $E$, i.e., $T$. This bootstrap procedure does not break down the information flow from $B$ and $C$ to $A$. For each bootstrap realization we estimate $S^{B}_{(r)}$ and $S^{C}_{(r)}$. Finally the p-value of $\mbox{TE}^{B \to A}>\mbox{TE}^{C \to A}=S^{C}$ is defined as
\begin{equation}
    p\mbox{-value} = \frac{\sum_{i=1}^q\delta(S^{B}_{(i)}>S^{B}_{(i)})}{q} 
\end{equation}
where $q$ is the number of independent bootstrap re-sampling and $\delta(x)=1$ if the assertion is true, zero otherwise.

\section{Finite size effects of the statistical tests}
\subsection{Testing for the significant TE}\label{sec:null}
In this section, we show that the analytical p-value converges to the bootstrap one for various numbers of the number of states, hence of the degrees of freedom. In particular, we test first the convergence when the null hypothesis is true, then we test the power in detecting when the null hypothesis is false.

\begin{figure}[tbh]
    \centering
    \includegraphics[width=0.45\columnwidth]{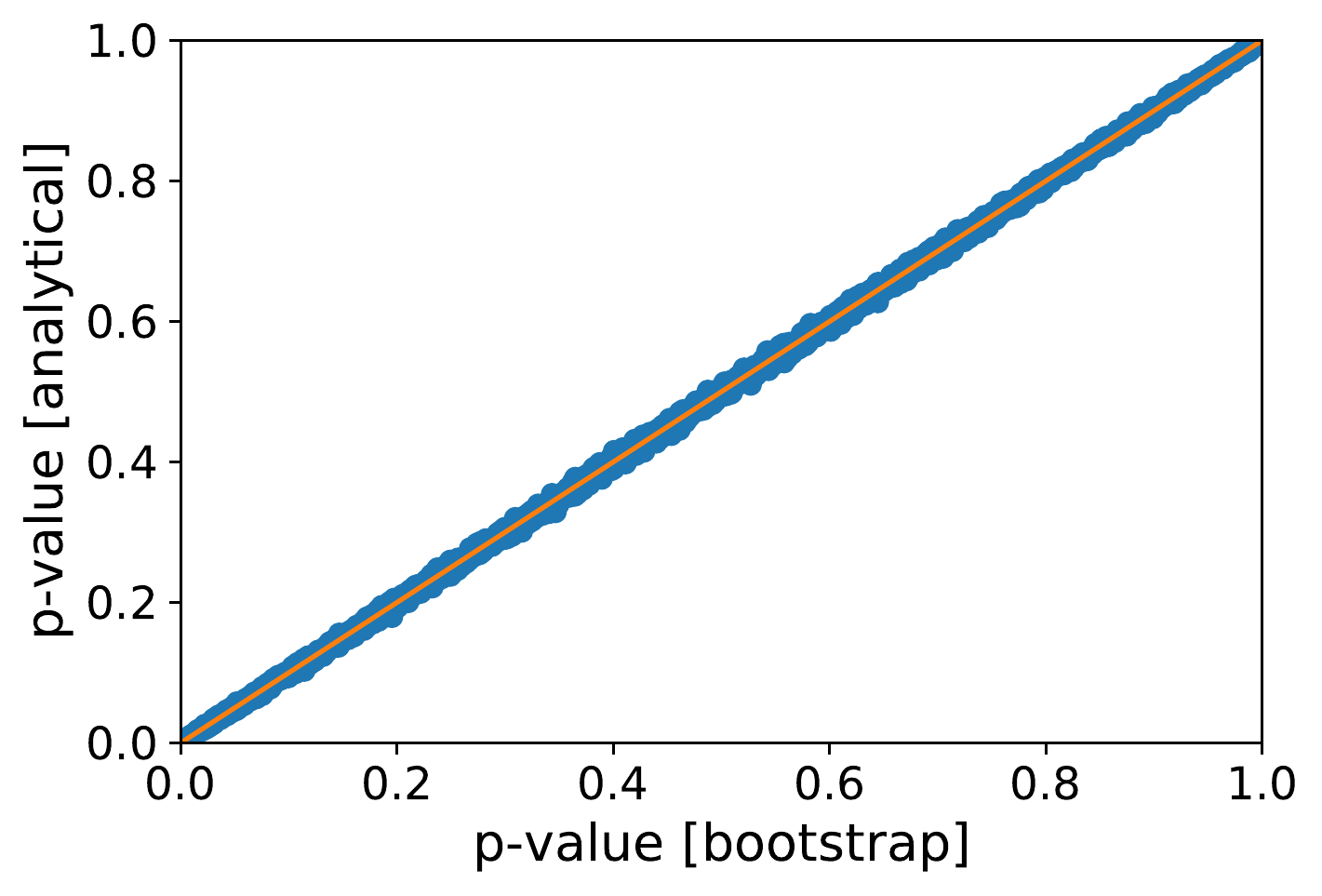}
    \includegraphics[width=0.45\columnwidth]{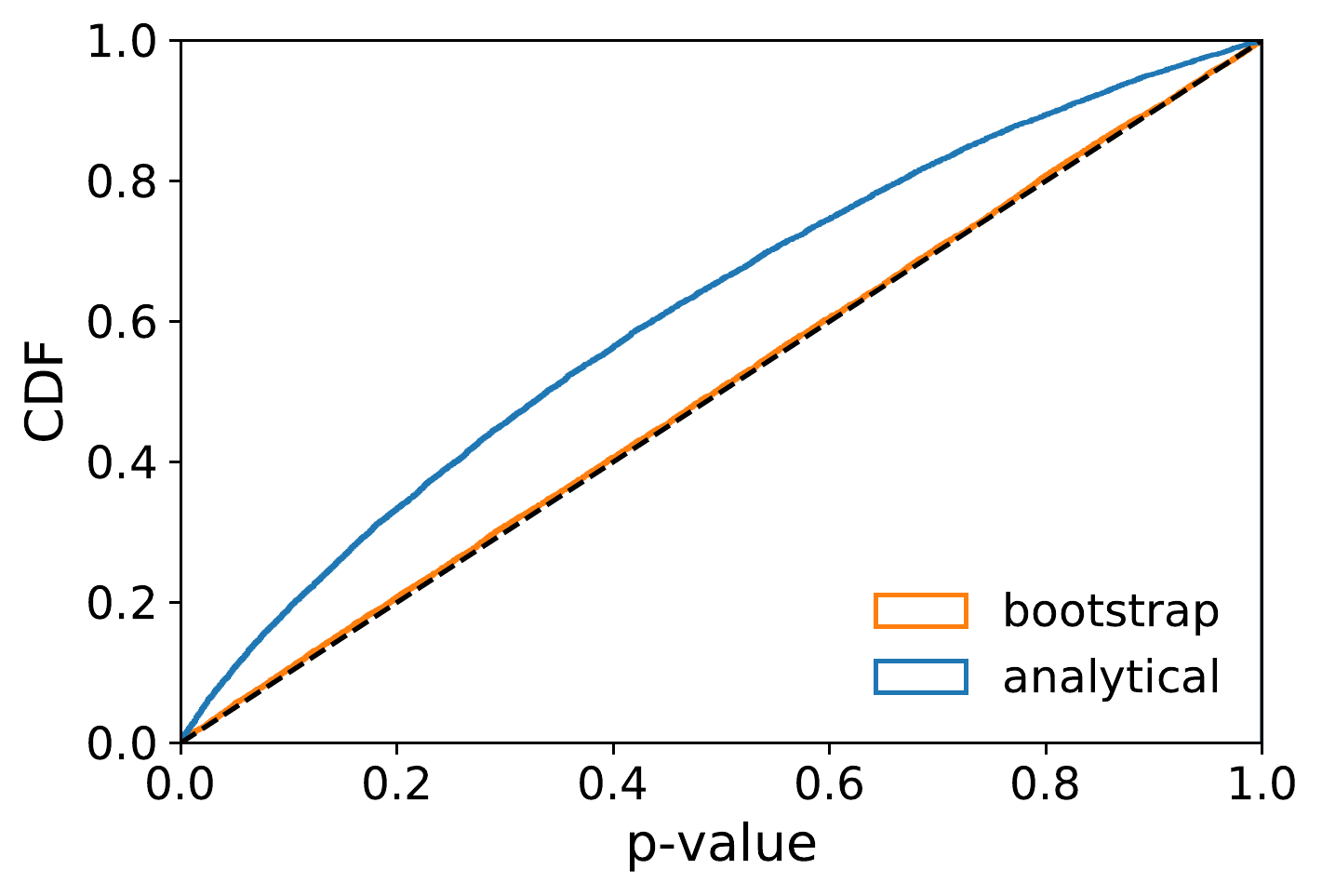}
    
    \includegraphics[width=0.45\columnwidth]{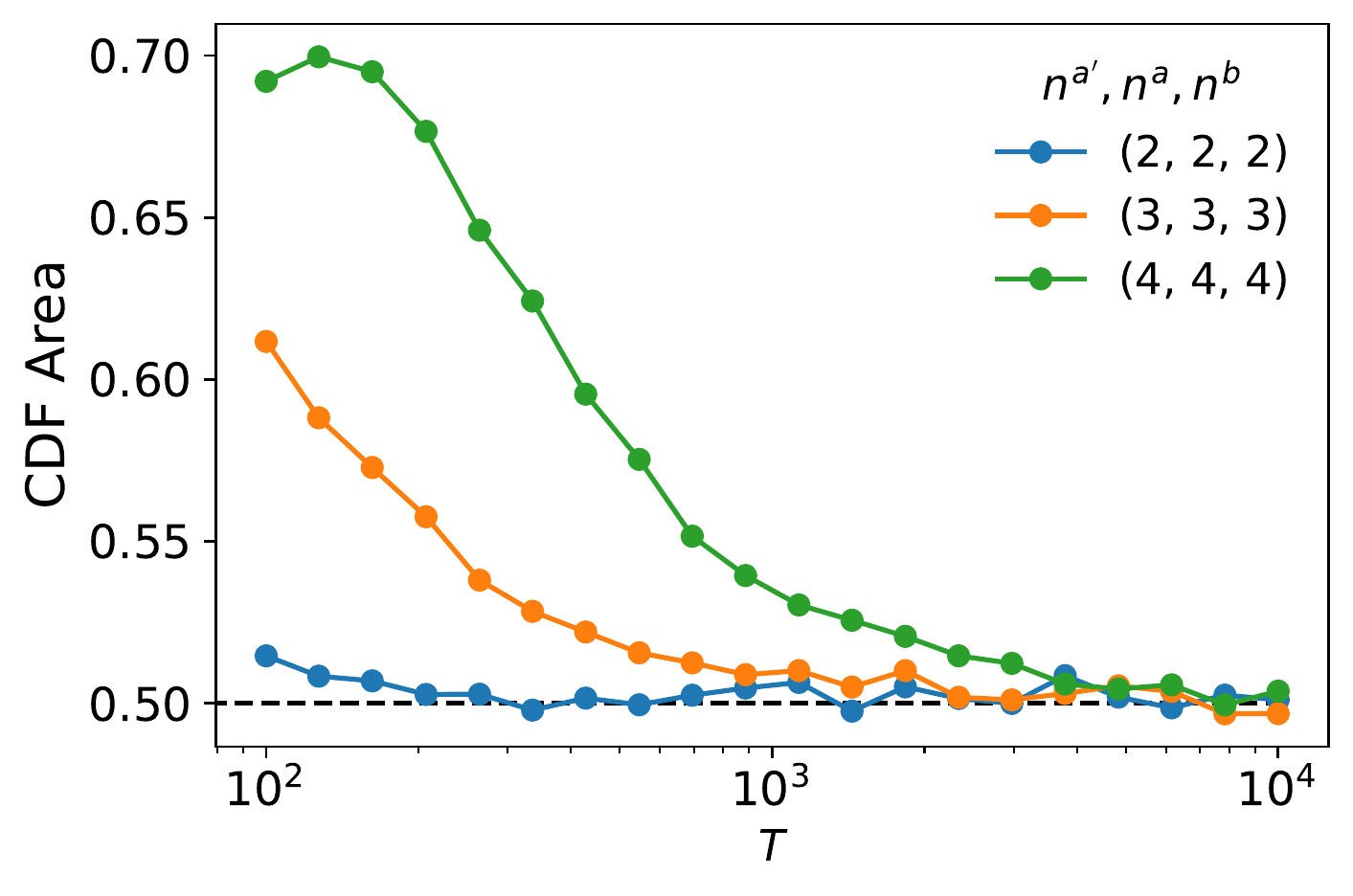}
    \includegraphics[width=0.45\columnwidth]{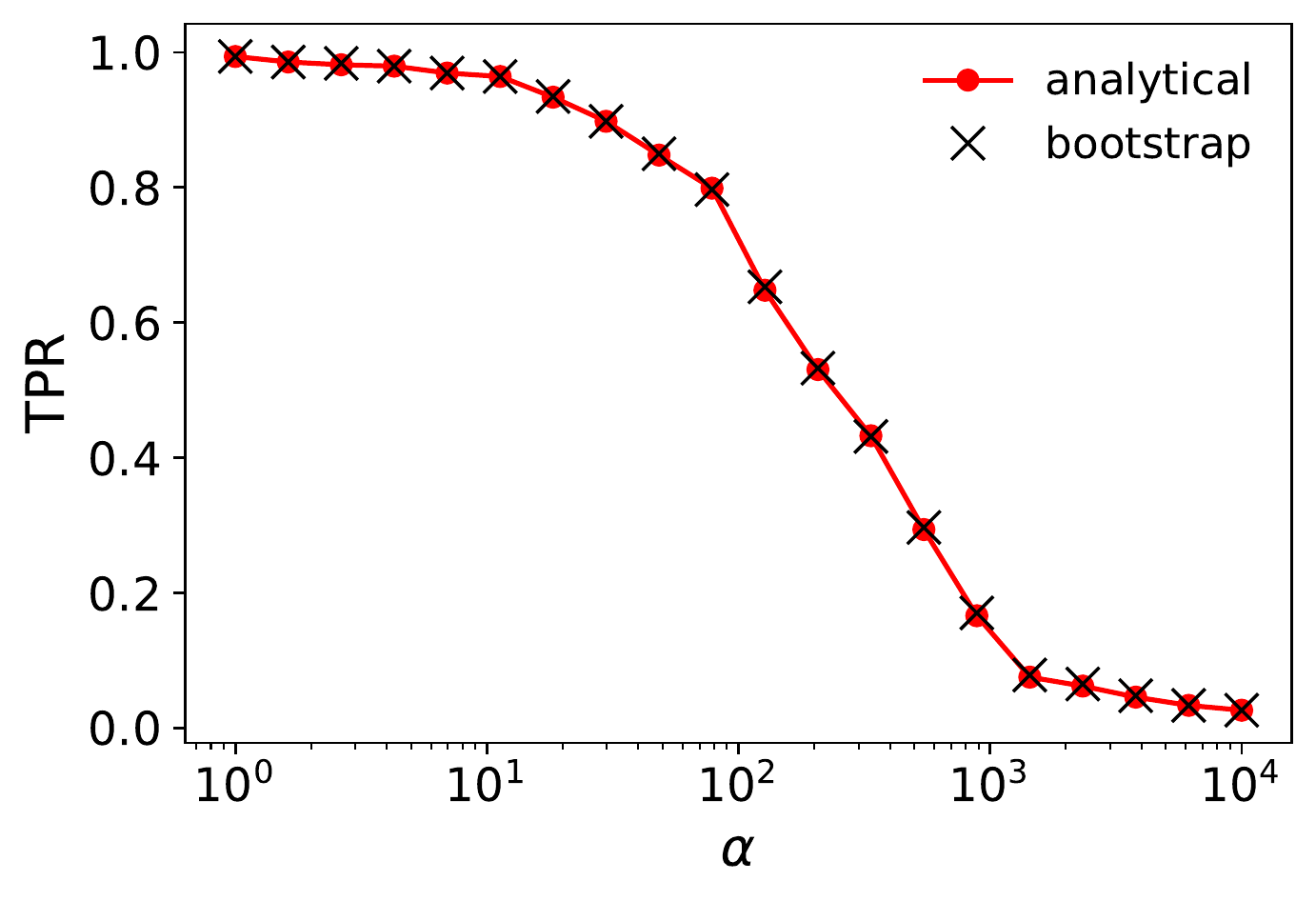}
    \caption{Tests for $\mbox{TE}^{B \to A}>0$: $a)$ comparison of bootstrap and analytical p-values for true null hypothesis with $T=10,000$ and all possible combinations of $(n^{A^+}n^A n^B) \in [2,4]$ repeated $100$ times; $b)$ cumulative distributions computed with $10,000$ independent realizations of the bootstrap and analytical p-values for $(n^{A^+}n^A n^B)=(3,3,3)$ and $T=100$, the unbiased model should lie on the bisector; $c)$ Area under the CDF of analytic p-value for different sample size $T$, the area of the unbiased model is $0.5$;  $d)$ comparison of the bootstrap an analytic true positive rate obtained with a p-value$<0.01$ for false null hypothesis benchmark, the simulations refers to $1,000$ simulation with $(n^{A^+}n^A n^B)=2,2,2$ and $T=10,000$, the parameter $\alpha$ controls the homogeneity of $\mathcal{P}(a',a,b)$.  In all the plots the bootstrap p-value is computed with $10,000$ bootstrap re-sampling.  }
    \label{fig:bech}
\end{figure}

The true null hypothesis benchmark is obtained by sampling $\mathcal{P}(a',a)$ and $P(b)$ from two Dirichlet distributions with parameters set to $\alpha$. Then a synthetic dataset of sample size $T$ is obtained sampling a set of that from $\mathcal{P}(a',a,b)=\mathcal{P}(a',a)\mathcal{P}(b)$. In the upper panel left of Fig.~\ref{fig:bech} we show that both the analytical and bootstrap p-values converge to the same value irrespective of $n^{A^+}$, $n^A$, and $n^B$. However, we must be aware that this is not totally true. In fact, when $n^{A^+},n^A,n^B$ are large and when the sample size is small,  the likelihood ratio test should not be applied due to slow convergence to the chi-squared distribution (see the upper right panel of Fig.~\ref{fig:bech}). In particular, a small $T$ leads to an under-estimation of the analytical p-values. In the lower-left panel of Fig.~\ref{fig:bech}, we show that small  $n^{A^+},n^A,n^B$ are much more robust to finite sample size effects than the larger ones.  

To test the false null hypothesis, the probabilities $\mathcal{P}(a',a,b)$ are sampled from a Dirichlet distribution with parameters set to $\alpha$. In this set of simulation, the effect of $\alpha$ is explored in order to test the power of the test: small values of $\alpha$ produce very heterogeneous probabilities, while when $\alpha>>1$,  all the probabilities become homogeneous, i.e.,  $\mathcal{P}(a',a,b) \to 1/(n^{A^+}n^A n^B)$. While the former is the ideal condition for every test, the latter is a difficult case. In the lower-right panel of Fig.~\ref{fig:bech}, we show the fraction of times the analytical and bootstrap p-values reach values smaller than $0.01$ over $1,000$ independent simulations. Note that both the asymptotic analytical and the bootstraps results agree.

\subsection{Comparing TEs}\label{sec:over}
Benchmarking the true null hypothesis, in this case, is much harder. To do that, we should solve a non-linear minimization problem. Specifically, by using a Semidefinite Quadratic Linear Programming (SQLP), we build a set of probability $\mathcal{P}(a',a,b,c)$ such that $\left( \mbox{TE}^{c \to a}-S^{c})^2+(\mbox{TE}^{b \to a}-S^{b}\right)^2$ computed on the real benchmark probabilities is minimal. Since SQLP depends on the initial conditions, we randomize them by starting the optimization procedure from a set of probability obtained from a Dirichlet distribution with parameter $\alpha=0.5$.

\begin{figure}[tbh]
    \centering
    \includegraphics[width=0.45\columnwidth]{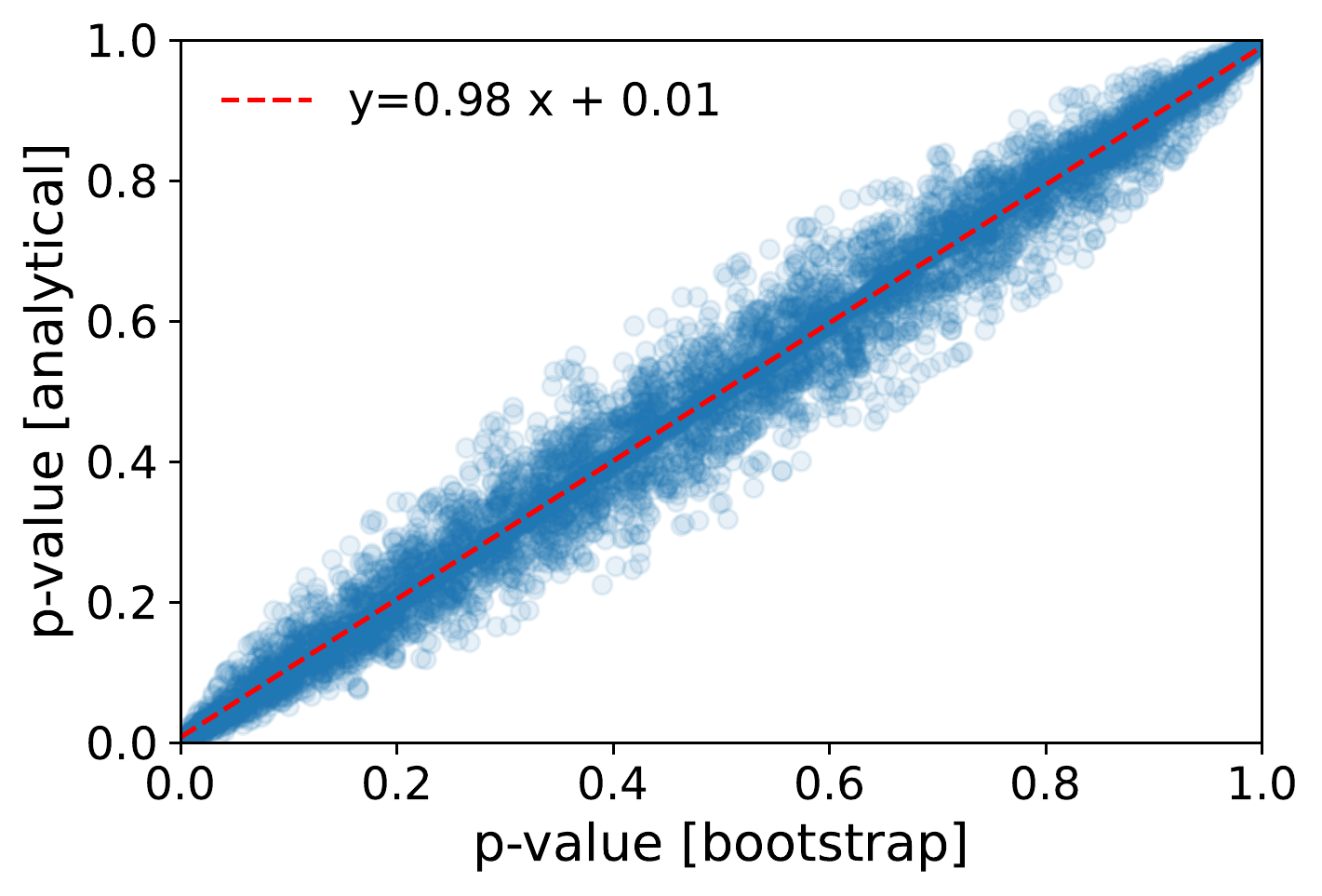}
    \includegraphics[width=0.45\columnwidth]{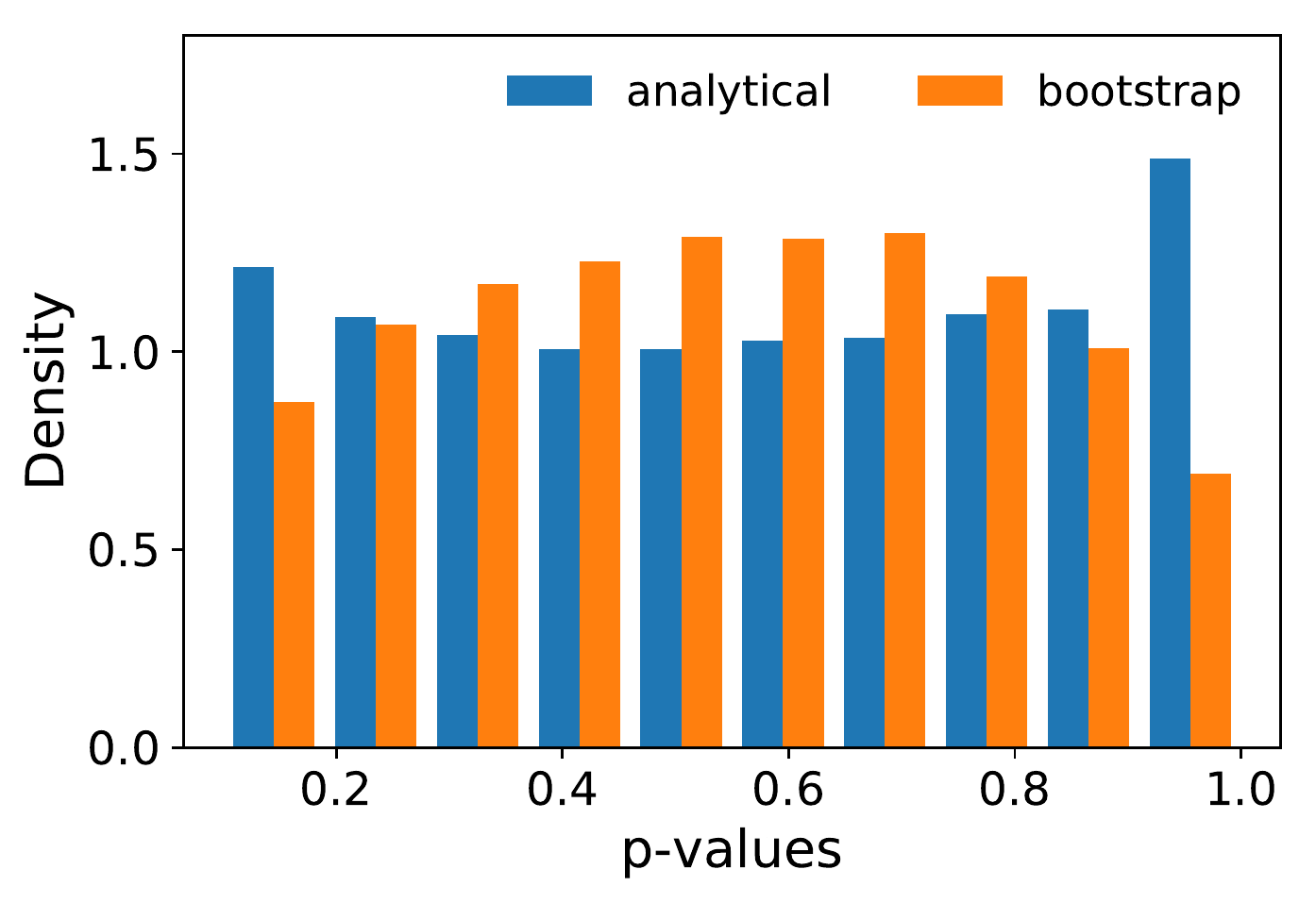}
    \includegraphics[width=0.45\columnwidth]{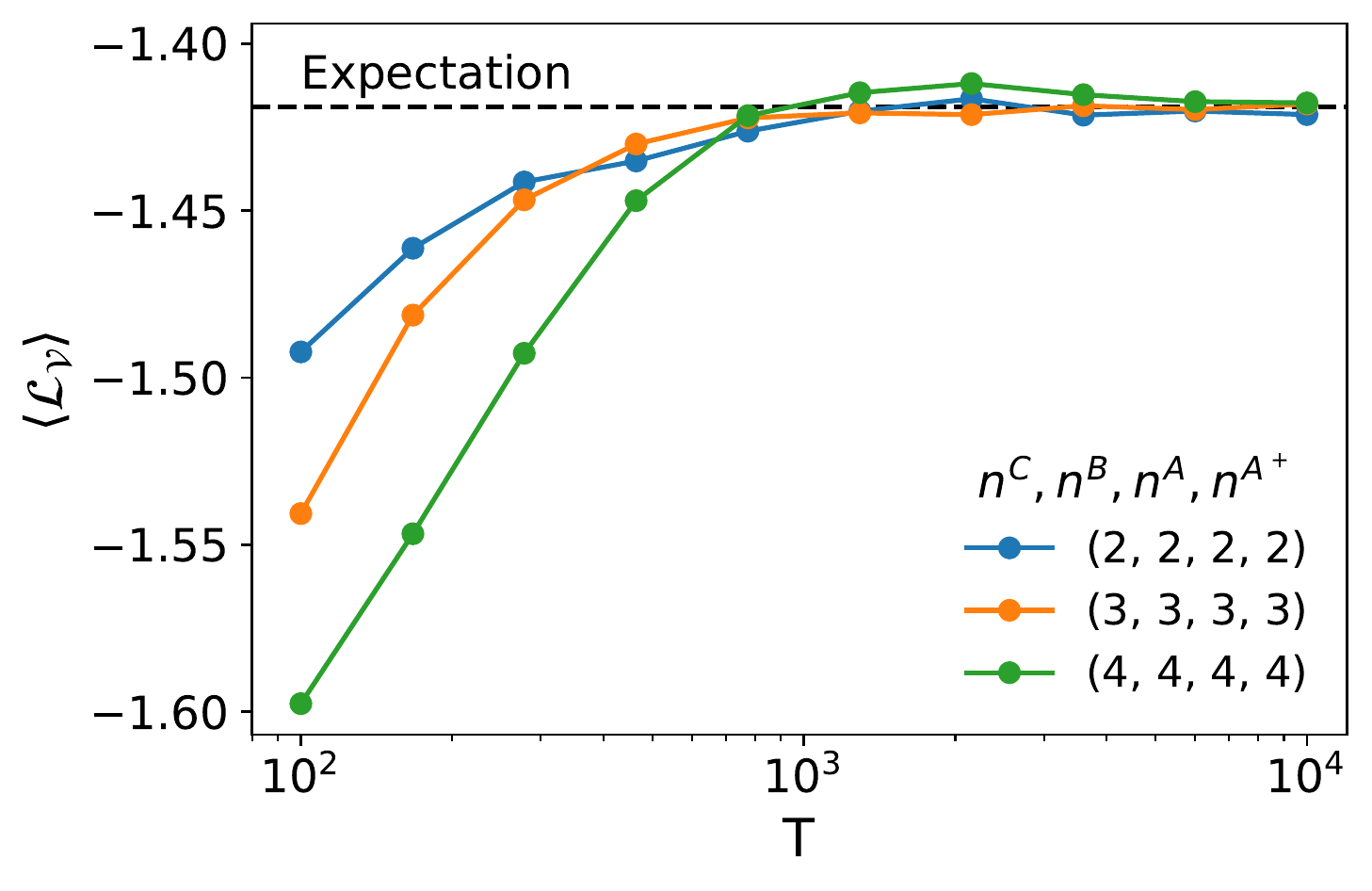}
    \includegraphics[width=0.45\columnwidth]{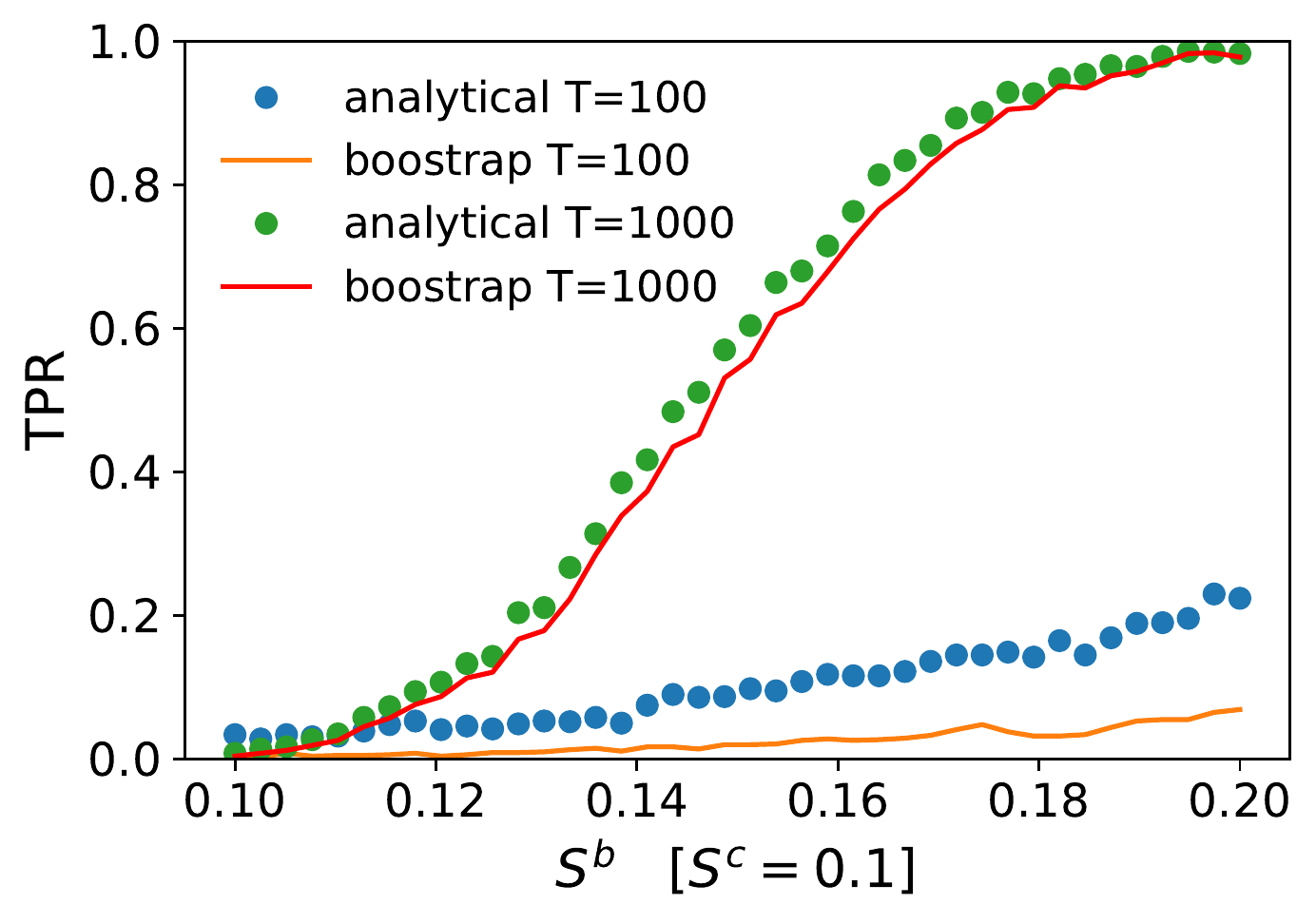}
    \caption{Benchmarks performance of the Vuong test. Upper left plot: scatter plot of the bootstrap p-values and the analytical ones for all possible combinations of $n^{A^+},n^{A},n^{B},n^{C} \in [2,4]$ for $T=10,000$ for $S^B=S^C=0.1$, the dotted line is a linear regression. Upper right plot: distribution of $10,000$ p-values obtained for $n^{A^+}=n^{A}=n^{B}=n^{C}=4$ and $T=100$ and  $S^B=S^C=0.1$, the expected distribution should be homogeneous. Lower left plot: average element-wise likelihood of the Vuong metric being distributed as a $\mathcal{N}(0,1)$ and $S^B=S^C=0.1$ while the dotted line is the likelihood corresponding to  $\mathcal{N}(0,1)$. Right lower plot: true positive rate computed over $1,000$ simulations for $n^{A^+}=n^{A}=n^{B}=n^{C}=4$ as a function of $S^B$ for different values of $T$.}
    \label{fig:vuongbench}
\end{figure}

In the upper panel of Fig.~\ref{fig:vuongbench} we show a scatter plot of the p-values obtained when the null hypothesis is true for a large sample size. Differently from Fig.~\ref{fig:bech}, this time bootstrap and analytical p-values show a lower agreement, although on average unbiased, as depicted in the linear regression reported in the plot. 

In the low sample size and large number of symbols regime (the upper right panel of Fig.~\ref{fig:vuongbench}), both p-values are biased but in a different way. The bootstrap p-values are too restrictive, while the analytical ones are too permissive. In fact, the expected p-value distribution when the null hypothesis is true should be a homogeneous distribution. In this case, since the Vuong test is two-sided, looking at the area of the CDF of p-values may be misleading. Therefore, to inspect the speed of convergence of the analytical p-values, we measure the average element-wise likelihood that the Vuong metric $\mathcal{V}$ is distributed as a $\mathcal{N}(0,1)$ (see the lower-left plot of Fig.~\ref{fig:vuongbench}). As expected, the bias is larger when the number of symbols is large and all the inspected cases converge to normality for $T>1,000$.

Finally, in the lower right panel of Fig.~\ref{fig:vuongbench}, we compare the true positive rate of the bootstrap and analytical p-values. The analytical p-values, as expected, show a higher recall than the bootstrap ones. Although this advantage is larger in the small sample size regime, it is worth stressing that it might be compensated by a lower control of the first type errors.

\section{Lagged Information transfer between asynchronous time series: limit order books}

We apply the analytical asymptotic p-values described above to measure the information flow between changes in the order books of different stocks. Specifically, for each stock, we consider the time series of mid-prices, defined as the instantaneous average between the best buy and sell quotes, and encode their changes with their signs. The inter-event time between two consecutive price variations is not constant, and two time series are not synchronous. Ref.~\cite{harre2015entropy} generalized TE to asynchronous time series by matching a given event of time series $A$  with the last known event of time series $B$. This type of matching has two consequences: i) one event of $B$ may be matched to multiple events of $A$; ii) some event of $B$ might be not matched to an event of $A$, especially if the typical event frequency is much larger for $B$ than for $A$.


\begin{figure}[tbh]
    \centering
    \includegraphics[width=0.95\columnwidth]{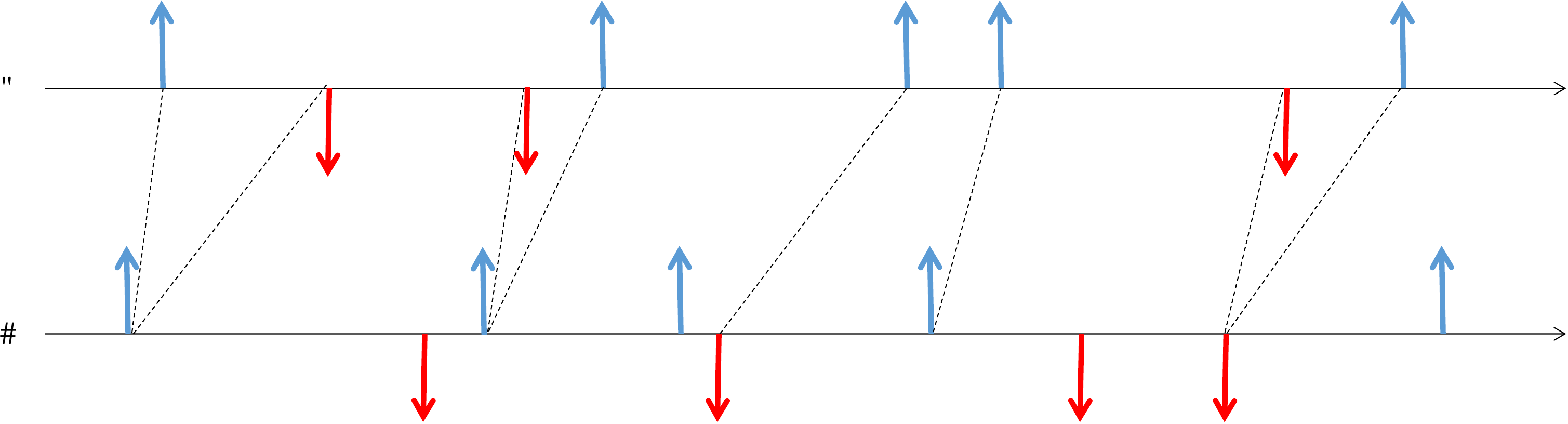}
    \includegraphics[width=0.80\columnwidth]{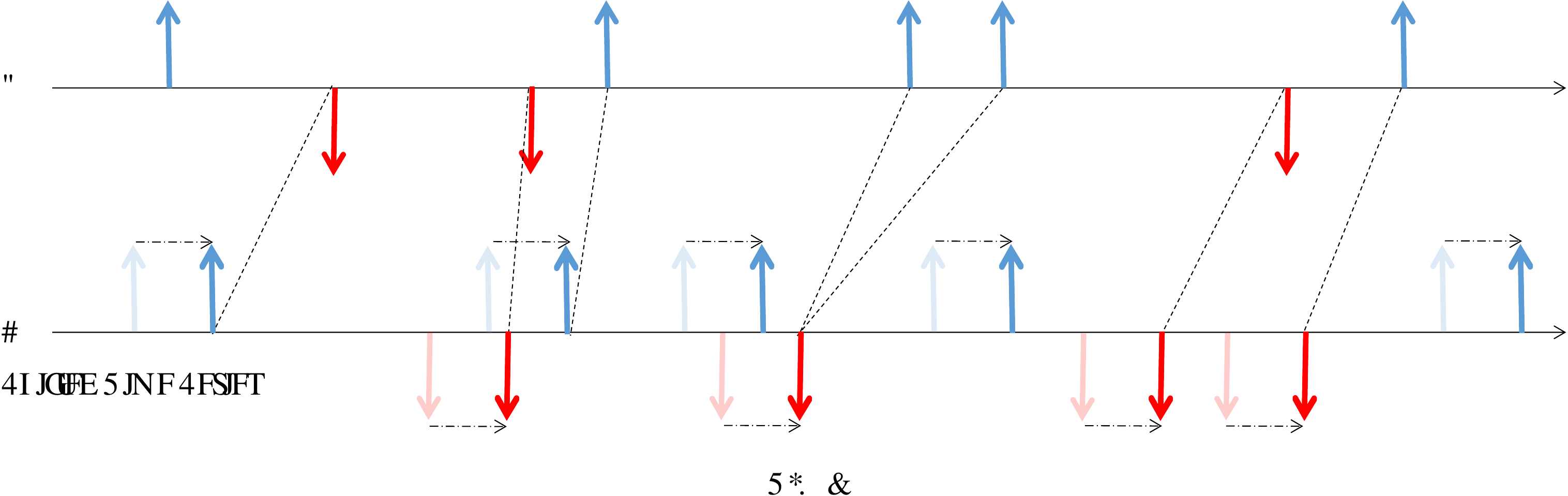}
    \caption{The upper plot shows an example of the backward matching among two asynchronous binary series. The lower plot shows the resulting match after a forward shift is applied to X.   }
    \label{fig:sync}
\end{figure}

It is interesting to observe that the reaction time is not immediate in real applications.
 Therefore, we expect that an event on $B$ might take some time to process and affect the future of $A$. To measure the amount of this transient, we apply a forward shift to the time series $A$ of $\delta t$ and apply the matching again. (Fig.~\ref{fig:sync}).
 \begin{figure}
    \centering
    \includegraphics[width=0.45\columnwidth]{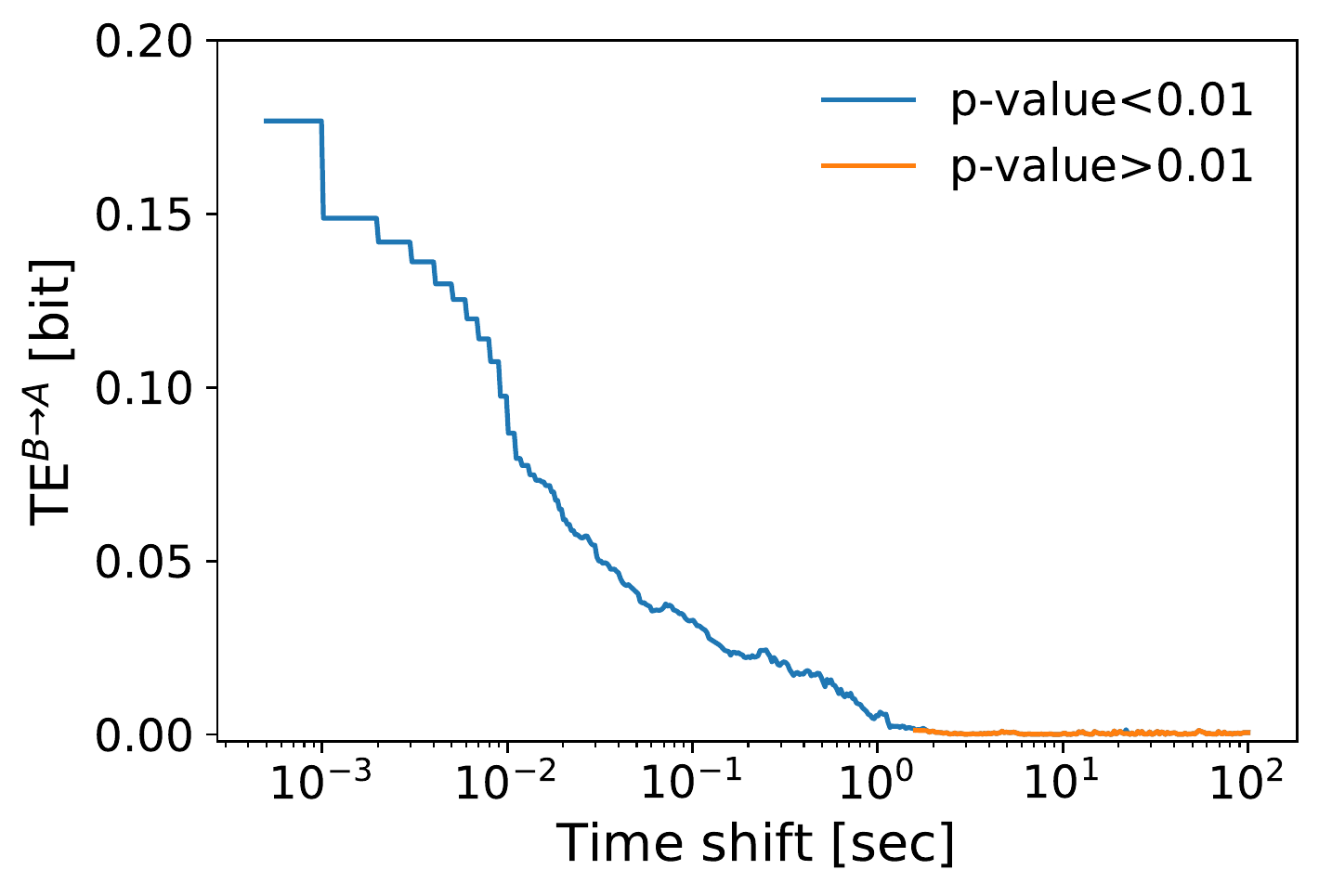}
    \includegraphics[width=0.45\columnwidth]{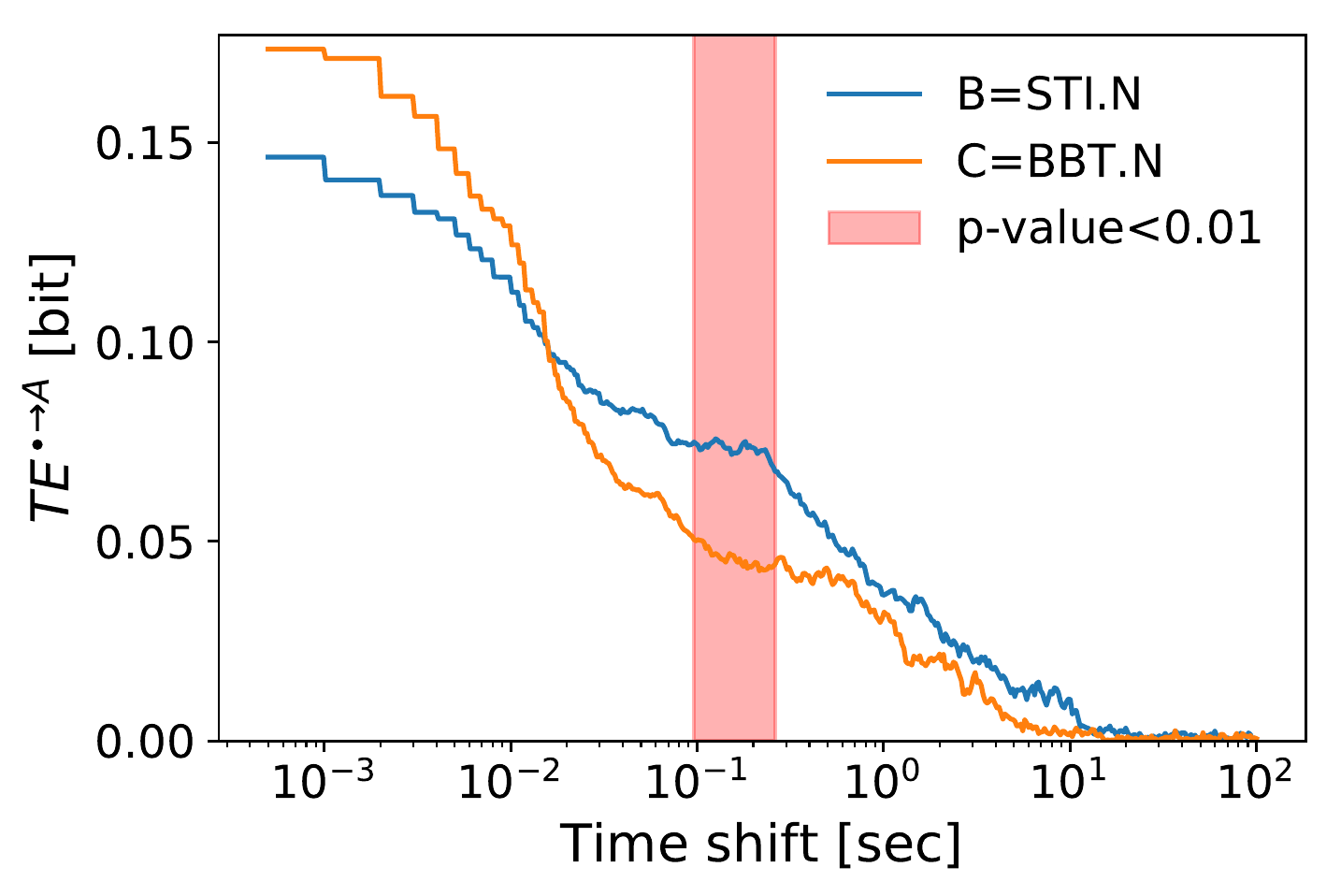}
    \caption{Left plot: TE from {\tt NEM} to {\tt ABX} as a function of the forward shift of {\tt ABX}; when the shift is larger than $1$ second,   TE is no more significant. The right plot: TE from {\tt STI} and {\tt BBT} to {\tt RF} as a function of the time shift of {\tt RF} computed over 3 days of data. The difference in TE is significant according a two-sided Vuong test only in the red band}
    \label{fig:oob}
\end{figure}

In the left plot of Fig.~\ref{fig:oob}, we show the TE from stock {\tt NEM} to stock {\tt ABX} computed on 2017-01-03, selecting only events between $09:35$ and $16:25$ (we removed 5 minutes at both ends of the trading day to reduce the effect of the auction phase). As expected, the TE drops down as the shift increases. By using the statistical test of sec.~\ref{sec:tepv} we can detect the maximum $\delta t\simeq 1s$ after which the binary price changes of  {\tt NEM} do not add any information to the predictability of those of {\tt ABX}. Note that this delay does not allow one to make useful price predictions in a speculative setting, as prices move too little in 1 second to beat transaction costs.
We emphasize here that the ability to perform this test on long time series without bootstrapping opens the way to large-scale studies of the timescales of TE and delays. 

In the right plot of Fig.~\ref{fig:oob} we apply the TE comparison test with time shifts to similar data: we compare the TE from \texttt{STI} and \texttt{BBT} to {\tt RF} as a function of the time shift $\delta t$. In this case, it seems that when $\delta t<0.01$s, {\tt BBT}  transfers more information to {\tt RF} than {\tt BBT}, whereas on a larger time scales ($\delta t>0.01$s) we observe the opposite. This, however, may simply be due to estimation noise. Indeed, the TE comparison test is only significant in the area highlighted in red at a $0.01$ threshold, whereas the comparison for $\delta t<0.01$ is not.

\subsection{Network inference}

The analytic p-values of sec.~\ref{sec:tepv} can be used to extract the lead-lag network between different stocks. Specifically, we considered all the order books of NYSE and five German exchanges on a single day. Given that the information flow between the European and American exchange must be penalized due to physical distance, we expect to observe higher link homophily within a continent.

We restricted our analysis to a single day (2017-01-03) and selected only the price variations during the daytime when all the exchanges were simultaneously opened (14:30-16:30 UTC). In order to avoid small sample size biases of the analytic p-values, reported in sec.~\ref{sec:null}, we removed all the stocks which have less than $1,000$ mid-price variations and we considered only binary variations, i.e, $P_{t+1}>P_{t}$ or $P_{t+1}<P_{t}$. After this data filtering, only $n=1,177$ stocks remain. Given that we are performing a multiple comparison test to control the number of false-positive selections, we adopted the Bonferroni correction~\cite{millersimultaneous} which prescribes reducing the null hypothesis threshold for each test to $q/n_t$ where $1-q$ is the confidence level at which we can guarantee that no false link is included in the inferred network ($q=0.01$) and $n_t=n(n+1)/2=693,253$ are the number of tests performed. We want to stress that for such a problem using a bootstrap p-value will be computation unfeasible due to the large number of tests performed and the low Bonferroni threshold.

\begin{figure}
    \centering
    \includegraphics[width=0.45\columnwidth]{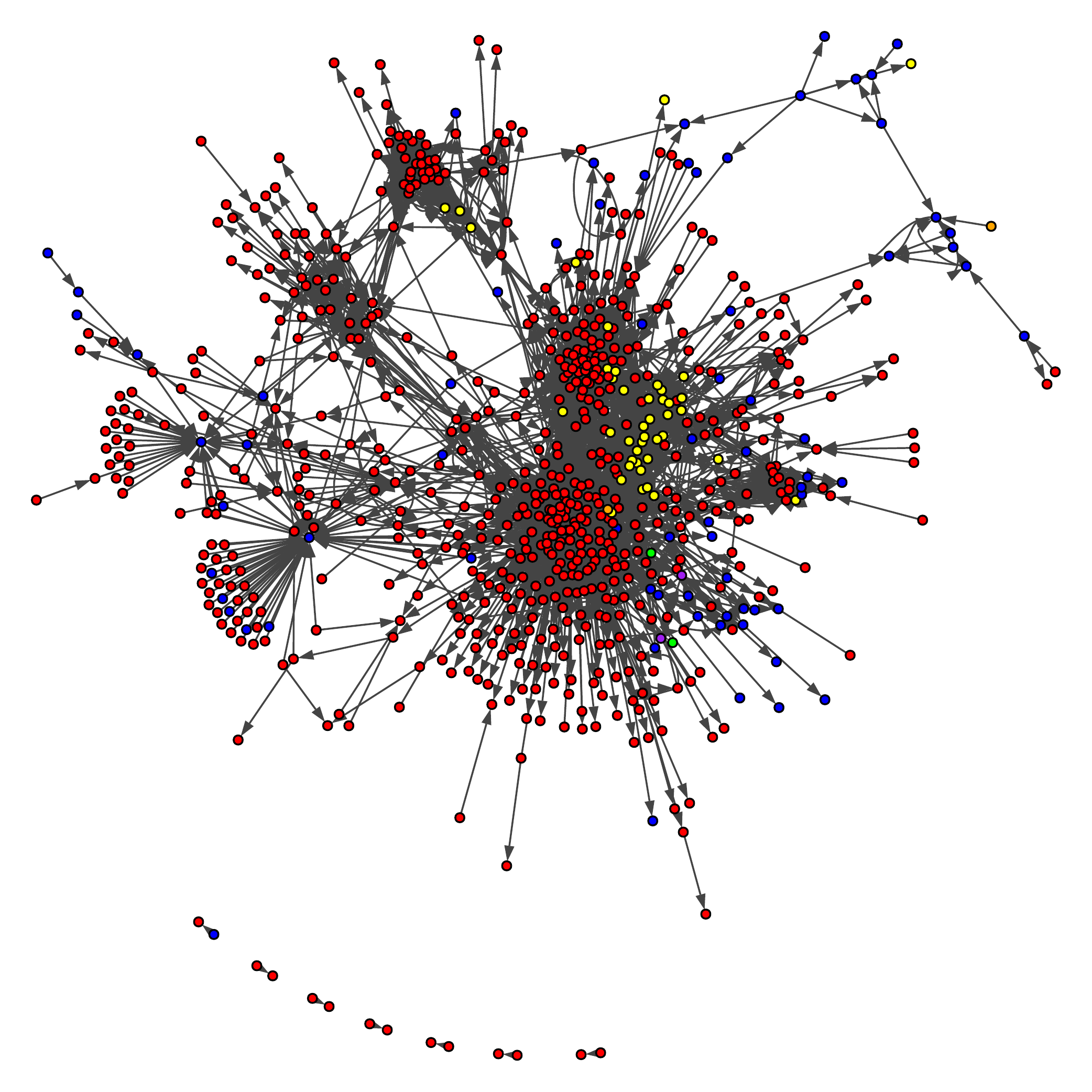}
    \includegraphics[width=0.45\columnwidth]{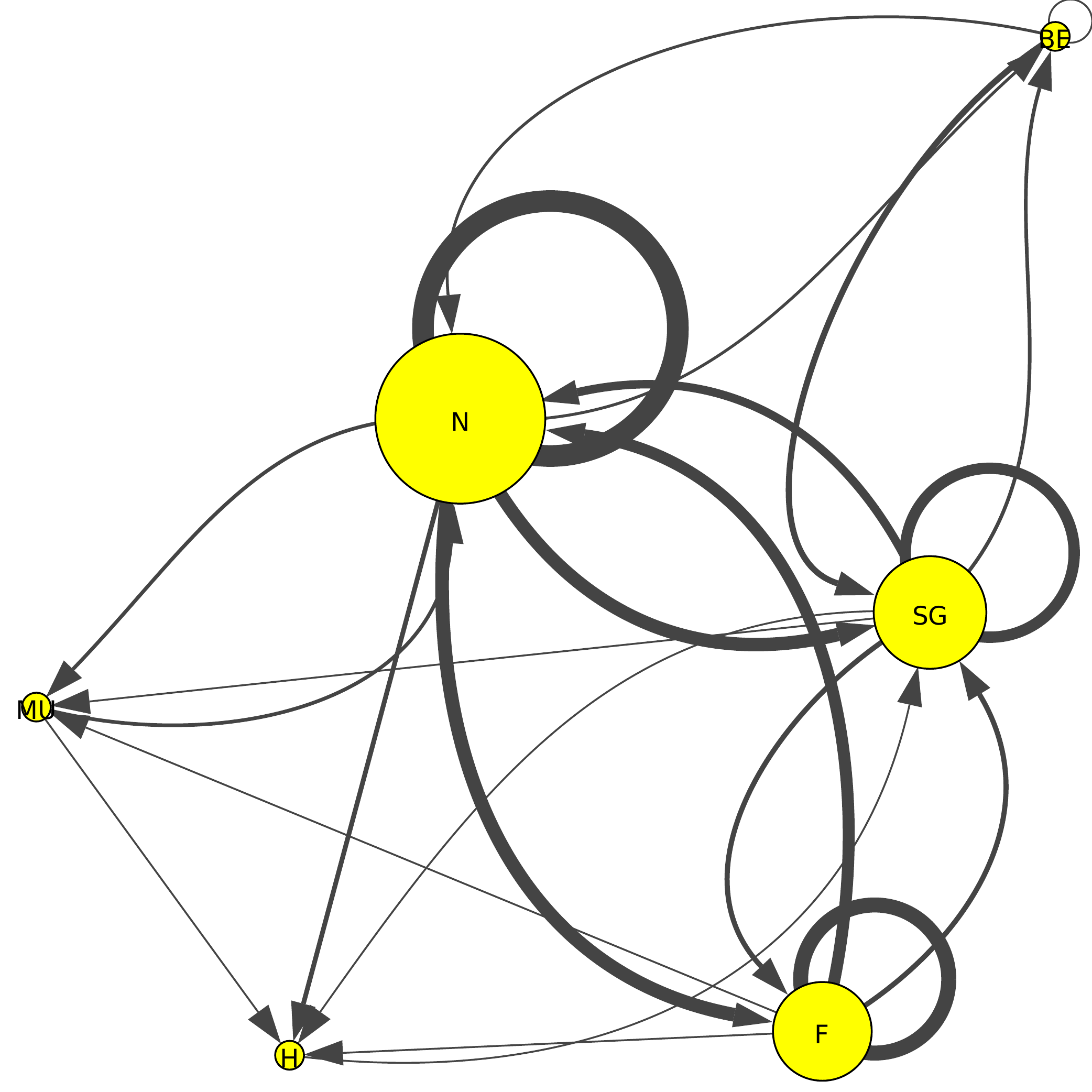}
    \caption{Left plot: Bonferroni network obtained retaining only the significant TE among each pair of companies of NYSE and 4 German stock exchanges; the color code indicates the stock exchanges. Right plot:  network representation of the information flow among the stock exchanges. The size of the nodes is proportional to the log of the exchange's size, and the links' width is proportional to the log of the number of links among the considered groups. }
    \label{fig:net}
\end{figure}

On the resulting statistically validated network \cite{tumminello2011statistically}, shown in Fig.~\ref{fig:net} left, only $0.3\%$ of links were retained, and $767$ nodes were not isolated. As we expected\cite{bongiorno2017core}, we observed a high assortativity coefficient~\cite{newman2010networks} by exchange, i.e., $0.56$, which means that the number of links inside the same exchange is larger than the one expected from a configuration model.

In the right panel of Fig.~\ref{fig:net}, we display a coarse-grained network, where all the nodes that belong to the same exchanges are grouped into a single node with a size proportional to the log of the number of nodes, and all the links are grouped with a width proportional to the log the number of links. From this analysis, one observes that the largest interconnection is between NYSE and Frankfurt stock exchanges, which are the most important stock exchanges in our dataset. 

\section{Conclusions}

Having asymptotic tests for the presence of Transfer Entropy and a comparison of TE makes it possible to infer causality networks of a large number of nodes with long time series, especially since the minimum number of points required is relatively small. Using a time shift is especially relevant in systems in which latency is important, which deserves further study. Finally, being able to compare the TE from two sets of predictors of a given time series opens many possibilities, such as pair-wise ranking of feature importance.

%
%

\bibliographystyle{plain}

\bibliography{references}

\end{document}